# Multizone sound field reproduction with direction-of-arrival-distribution-based regularization and its application to binaural-centered mode-matching

Ryo MATSUDA[1,2]; Yuki KAWASAKI[1]; Makoto OTANI[1]

[1] Graduate School of Engineering, Kyoto University, Japan

[2] Research and Development Division, Yamaha Corporation, Japan

**ABSTRACT**

In higher-order Ambisonics, a framework for sound field reproduction, secondary-source driving signals are generally obtained by regularized mode matching. The authors have proposed a regularization technique based on direction-of-arrival (DoA) distribution of wavefronts in the primary sound field. Such DoA-distribution-based regularization enables a suppression of excessively large driving signal gains for secondary sources that are in the directions far from the primary source direction. This improves the reproduction accuracy at regions away from the reproduction center. First, this study applies the DoA-distribution-based regularization to a multizone sound field reproduction based on the addition theorem. Furthermore, the regularized multizone sound field reproduction is extended to a binaural-centered mode matching (BCMM), which produces two reproduction points, one at each ear, to avoid a degraded reproduction accuracy due to a shrinking sweet spot at higher frequencies. Free-field and binaural simulations were numerically performed to examine the effectiveness of the DoA-distribution-based regularization on the multizone sound field reproduction and the BCMM.

## 1. INTRODUCTION

Sound field reproduction is a technique for physically reproducing a real or virtually simulated sound field (1, 2). The sound field is generally recorded using a microphone array (3, 4) and reproduced using a loudspeaker array (5, 6, 7) or headphones with head-related transfer functions (8, 9). Higher-Order Ambisonics (HOA) is a framework for recording, analyzing, and reproducing a sound field by expanding a sound field in terms of spherical harmonic functions. However, an infinite order of spherical harmonic expansion is truncated depending on the number of microphones and loudspeakers in practice. The size of the recorded and reproduced sound field, which is so called sweet spot, is limited to a spherical region at a certain distance from the origin of the spherical harmonic expansion due to the nature of the radial function when the infinite order is truncated. In general, sweet spot size is inversely proportional to frequency.

When reproducing a sound field using a loudspeaker array in HOA, mode matching is generally used, which is a method of determining the loudspeaker driving signals so that the expansion coefficients of the primary sound field and those of the reproduced sound field synthesized by the loudspeaker array are matched in the least-squares sense. In mode matching, a regularization technique based on direction-of-arrival (DoA) distribution of wavefronts in the primary sound field has proposed (10). It has been shown that using DoA-distribution-based regularization suppresses unwanted output of secondary sources, especially in the opposite direction to the primary source, and improves the accuracy of reproduction at positions away from the sweet spot.

This study applies DoA-distribution-based regularization to multizone sound field reproduction and evaluates its accuracy. Furthermore, the DoA-distribution-based regularization is also applied to binaural-centered mode matching (BCMM) (11), which generates two sweet spots at the ear positions of the listener.

## 2. Multizone reproduction with DoA-distribution-based regularization

### 2.1 Formularization

The loudspeaker driving signals for multizone reproduction is expressed as,

$$\boldsymbol{d} = (\boldsymbol{C}^{\mathrm{H}}\boldsymbol{C} + \lambda \boldsymbol{I})^{-1}\boldsymbol{C}^{\mathrm{H}}\boldsymbol{b}, \tag{1}$$

where $\boldsymbol{d} \in \mathbb{C}^{L \times 1}$ is the vector consists of loudspeaker driving signals; $\boldsymbol{C} \in \mathbb{C}^{Q(N_q+1)^2 \times L}$ consists of the spherical harmonic coefficients of the transfer function for the loudspeakers corresponding to the expansion around each local origin $O^{(q)}(q = 1, \dots, Q)$; $\boldsymbol{b} \in \mathbb{C}^{Q(N_q+1)^2 \times 1}$ is the vector containing the spherical harmonic coefficients of the primary sound field corresponding to $O^{(q)}$; $L$ is the number of the loudspeakers; $N_q$ is the truncation order for each zone; $Q$ is the number of multizone; $\lambda$ is the regularization parameter; $\boldsymbol{I} \in \mathbb{R}^{L \times L}$ is an identity matrix (5, 6).

DoA-distribution-based regularization, proposed by Kawasaki *et al.* (9), is applied to mode matching for multizone reproduction, expressed as,

$$\boldsymbol{d} = (\boldsymbol{C}^{\mathrm{H}}\boldsymbol{C} + \lambda \boldsymbol{\Sigma})^{-1}\boldsymbol{C}^{\mathrm{H}}\boldsymbol{b}, \tag{2}$$

where $\boldsymbol{\Sigma} \in \mathbb{R}^{L \times L}$ is a diagonal matrix given by $[\boldsymbol{\Sigma}]_{l,l} = 1/\|\mu(\boldsymbol{r}_l, k)\|$. Here, $\mu(\boldsymbol{r}_l, k)$ represents the weights of the spherical wave arriving from each direction on the sphere, known as the single-layer potential (1),

$$\mu(\boldsymbol{r}_l, k) = \sum_{n=0}^{N} \sum_{m=-n}^{n} \frac{\mathrm{j} b_n^m(k)}{k R_l^2 h_n^{(2)}(k R_l)} Y_n^m(\theta_l, \phi_l), \tag{3}$$

where $k$ is the wavenumber; $h_n^{(2)}(\cdot)$ is the spherical Hankel function of the second kind; $Y_n^m(\cdot)$ is the spherical harmonic function of n-th order and m-th degree; $\boldsymbol{r}_l(R_l, \theta_l, \phi_l)$ is the position of point source. Since the weights are relative value, the norm $\|\mu(\boldsymbol{r}_l, k)\|$ is normalized by the maximum value over all the loudspeakers. This weighting suppresses the output of loudspeakers located on the opposite side of the primary source.

### 2.2 Numerical simulation

A numerical simulation was conducted to compare mode-matching methods with two regularizations; one is the DoA-distribution-based regularization (**Proposed**) in Eq. (2), and the other is normal regularization (**Conventional**) in Eq. (1).

The loudspeakers were assumed as point sources, and 121 loudspeakers were placed on a sphere with a radius of 1.5 m whose angular coordinates were determined by spherical *t*-design (12). The truncation order $N_q$ in Eqs. (1) and (2) was set to 6, and $N$ in Eq. (3) was also set to 6. Two reproduced zones were centered at $(0.5, 0, 0)$ and $(-0.5, 0, 0)$. The simulation was performed for a single frequency, with the primary source being a point source with a frequency of 1 kHz in a free field.

The normalized reproduction error (NRE) (7) expressed as,

$$\mathrm{NRE}(\boldsymbol{r}, k) = 10 \log_{10} \frac{\int_\Omega |p_{\mathrm{rep}}(\boldsymbol{r}, k) - p_{\mathrm{des}}(\boldsymbol{r}, k)|^2 d\Omega}{\int_\Omega |p_{\mathrm{des}}(\boldsymbol{r}, k)|^2 d\Omega} \tag{4}$$

was calculated over spherical regions centered at each local origin of zone with a radius $R = 0.8N/k$, $N/k$, and $1.2N/k$ for each direction of primary source. The spherical regions are discretized as an orthogonal grid in 0.01 m intervals. The regularization parameter $\lambda$ in Eqs. (1) and (2) was set to the maximum eigenvalue of $\boldsymbol{C}^{\mathrm{H}}\boldsymbol{C} \times 10^{-3}$.

Figure 1 illustrates the results of NRE for the zone centered at (0.5,0,0). Results for only one zone are demonstrated because the centers of two zones are symmetric. The figure reveals that the proposed method has a smaller NRE when $R$ is greater than the radius determined by $N = kR$, which defines the sweet spot.

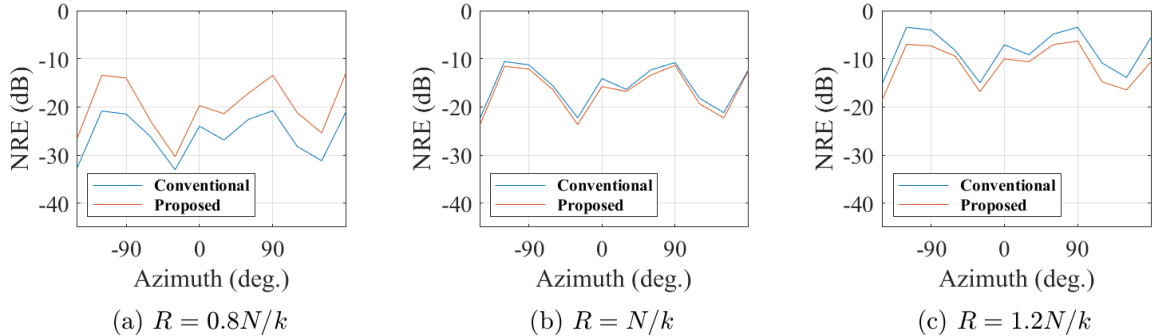

(a) $R = 0.8N/k$   (b) $R = N/k$   (c) $R = 1.2N/k$

Figure 1 – The results of NRE for each direction of primary source for zone centered at (0.5,0,0).

## 3. BCMM with DoA-distribution-based regularization

The DoA-distribution-based regularization was applied to BCMM (11). Reproduced binaural signals were simulated under the same conditions as in Sec. 2.2, except that the center of two zone were set to (0,0.0705,0) and (0,−0.0705,0). Reproduction performances were compared among the BCMM with DoA-distribution-based regularization (**BCMM w/ DoA**), the BCMM with conventional regularization (**BCMM w/ conv**) and the conventional mode-matching in global region (**MM**). The transfer functions from each loudspeaker or primary source to the ears of the dummy head were numerically calculated using the boundary element method (13).

Figure 2 demonstrates the normalized error (NE) of reproduced signals. Results only at the left ear are shown because the results are almost symmetrical between both the ears. **BCMM w/ DoA** has smaller NE than the other methods, especially in the high frequency range. As shown in the free-field simulation results in Sec. 2.2, DoA-distribution-based regularization suppressed the output of the loudspeaker located on the opposite side of the primary sound source and, therefore, reduces the error outside the sweet spot defined by $N = kr$, resulting in smaller errors in the binaural signals.

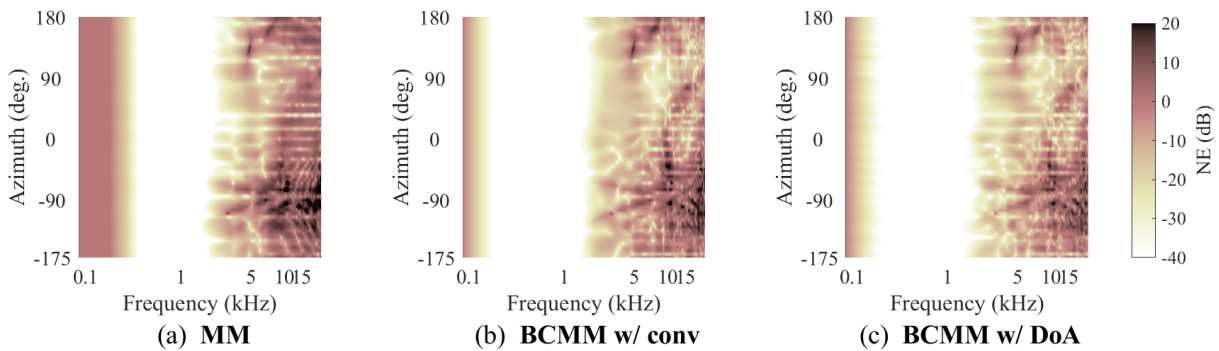

(a) **MM**   (b) **BCMM w/ conv**   (c) **BCMM w/ DoA**

Figure 2 – The results of NE distribution of reproduced binaural signals.

## 4. Conclusion

The DoA-distribution-based regularization was applied to multizone reproduction and BCMM. Numerical simulations showed that the DoA-distribution-based regularization achieves smaller reproduction error in the region slightly outside the sweet spot; it also leads to smaller error in the reproduced binaural signal when it is applied to BCMM.

**ACKNOWLEDGEMENTS**

This work was partly supported by grants-in-aid from JSPS, Japan (Grant Nos. 19H04153 and 19H04145).

**REFERENCES**

1. Poletti MA. Three-dimensional surround sound systems based on spherical harmonics. J. Audio Eng. Soc.; 2005; 53(11): 1004-1025.
2. Ward DB, Abhayapala TD. Reproduction of a plane-wave sound field using an array of loudspeakers. IEEE Trans. Speech Audio Process. 2001; 9(6): 697-707.
3. Rafaely B. Analysis and design of spherical microphone arrays. IEEE Trans. Speech Audio Process. 2004; 13(1): 135-143.
4. Samarasinghe P, Abhayapala T, Poletti M. Wavefield analysis over large areas using distributed higher order microphones. IEEE/ACM Trans. Audio Speech Lang. Process. 2014; 22(3): 647-658.
5. Wu YJ, Abhayapala TD. Spatial multizone soundfield reproduction: Theory and design. IEEE Trans. Audio Speech Lang. Process. 2010; 19(6): 1711-1720.
6. Zhang W, Abhayapala TD, Betlehem T, Fazi FM. Analysis and control of multi-zone sound field reproduction using modal-domain approach. J. Acoust. Soc. Am. 2016; 140(3): 2134-2144.
7. Ueno N, Koyama S, Saruwatari H. Three-dimensional sound field reproduction based on weighted mode-matching method. IEEE/ACM Trans. Audio Speech Lang. Process. 2019; 27(12): 1852-1867.
8. Noisternig M, Musil T, Sontacchi A, Holdrich R. 3D binaural sound reproduction using a virtual ambisonic approach. Proc IEEE International Symposium on Virtual Environments Human-Computer Interfaces and Measurement Systems 2003; 27-29 July 2003; Lugano, Switzerland 2003. p. 174-178.
9. Otani M, Shigetani H, Mitsuishi M, Matsuda R. Binaural Ambisonics: Its optimization and applications for auralization. Acoust. Sci. Tech. 2020; 41(1):142-150.
10. Kawasaki Y, Otani M. Mode-matching with regularization considering direction of arrival distribution of wavefronts in primary field. Proc. Spring Meet. Acoust. Soc. Jpn; March 2022; Japan.
11. Matsuda R, Otani M. Binaural-centered mode-matching method for enhanced reproduction accuracy at listener's both ears in sound field reproduction. J. Acoust. Soc. Am. 2021; 150(5): 3838-3851.
12. Tanaka T, Otani M. Modeling isotropic sound fields with spherical designs. Proc. Spring Meet. Acoust. Soc. Jpn; March 2022; Japan.
13. Otani M, Ise S. Fast calculation system specialized for head-related transfer function based on boundary element method. J. Acoust. Soc. Am. 2006; 119(5): 2589-2598.